\documentclass[%
 reprint,
superscriptaddress,
 amsmath,amssymb,
 aps,
prb,
]{revtex4-2}

\usepackage{amsmath}
\usepackage{graphicx}
\usepackage{dcolumn}
\usepackage{bm}
\usepackage[utf8]{inputenc}
\usepackage{mathptmx}
\usepackage{dutchcal}
\usepackage{pzccal}
\usepackage{color,soul}
\usepackage{hyperref}
\usepackage{bbold}
\usepackage{amssymb}
\usepackage{amsfonts}
\usepackage{textcomp}
\usepackage{eucal}

\DeclareMathAlphabet\mathbfcal{OMS}{cmsy}{b}{n}
\DeclareMathAlphabet\mathcalI{OMS}{cmsy}{m}{n}

\begin{document}


\title{{\it Ab-initio} transport theory for the intrinsic spin Hall effect applied to 5$d$ metals}

\author{Akash Bajaj}
\thanks{These two authors contributed equally}
 \affiliation{School of Physics and CRANN Institute, Trinity College, Dublin 2, Ireland}
 
\author{Reena Gupta}%
\thanks{These two authors contributed equally}
 \affiliation{School of Physics and CRANN Institute, Trinity College, Dublin 2, Ireland}
 
\author{Ilya V. Tokatly}
 \affiliation{Nano-Bio Spectroscopy Group and European Theoretical Spectroscopy Facility (ETSF), Departamento de Polímeros y Materiales Avanzados: Física, Química y Tecnología, Universidad del País Vasco (UPV/EHU), Av. Tolosa 72, 20018 San Sebastián, Spain}
 \affiliation{Donostia International Physics Center (DIPC), 20018 Donostia-San Sebastián, Spain}
 \affiliation{IKERBASQUE, Basque Foundation for Science, 48009 Bilbao, Spain}
\author{Stefano Sanvito}
 \affiliation{School of Physics and CRANN Institute, Trinity College, Dublin 2, Ireland}
\author{Andrea Droghetti}
 \email{andrea.droghetti@tcd.ie}
 \affiliation{School of Physics and CRANN Institute, Trinity College, Dublin 2, Ireland}




\begin{abstract}
We describe how the spin Hall effect (SHE) can be studied from {\it ab-initio} by combining 
density functional theory with the non-equilibrium Green's functions technique for quantum 
transport into the so-called DFT+NEGF method. After laying down our theoretical approach,
in particular discussing how to compute charge and spin bond currents, DFT+NEGF calculations 
are carried out for ideal clean systems. In these the transport is ballistic and the linear 
response limit is met. The SHE emerges in a central region attached to two leads when we apply 
a bias voltage so that electrons are accelerated by a uniform electric field. As a result, we 
obtain a finite spin-Hall current and, by performing a scaling analysis with respect to the 
system size, we estimate the ``ballistic'' spin Hall conductivity (SHC). We consider $5d$ 
metals with fcc and bcc crystal structures, finding that the SHC exhibits a rough qualitative
dependence on the $d$-band filling, and comment on these results in relation to existing 
literature. Finally, within the same DFT+NEGF approach, we also predict the appearance of 
a current-induced spin dipole moment inside the materials' unit cell and estimate its magnitude. 
\end{abstract}

\maketitle



\section{\label{sec:level1}Introduction}
The spin-Hall effect (SHE) in non-magnetic conductive materials refers to the generation 
of a spin current, transverse to the flowing charge current \cite{sinova2015spin,inoue2005taking}. 
It arises from the spin-orbit coupling (SOC) and appears as a spin accumulation at the samples' 
boundaries \cite{kato2004observation,sih2005}. The SHE was first predicted over 50 years 
ago \cite{dyakonov1971current,d1971possibility} and later rediscovered and theoretically 
studied in both metals and semiconductors \cite{ hirsch1999spin, zhang2000spin, murakami2003dissipationless, sinova2004universal,engel2005,raimondi2005, dimitrova2005}. 
It was then experimentally measured about two decades ago \cite{kato2004observation, sih2005,wunderlich2005experimental, stern2006, valenzuela2006direct}, 
ushering a large interest for spintronics-based device applications \cite{jungwirth2012spin,wang2013low}.

The most studied material systems for the SHE are heavy metals with large SOC, such as Pt and 
W~\cite{hoffmann2013spin, kimura2007room, Vila2007,liu2011spin,liu2012spin, pai2012spin, nakayama2013spin, Wang2014SHA,li2019materials,
zhu2021maximizing}. The common parameters used for quantifying the strength of the SHE are the 
spin-Hall conductivity (SHC) and the spin-Hall angle (SHA)  \cite{sinova2015spin, mosendz2010quantifying}, the latter being conventionally defined as the ratio of the SHC 
to the longitudinal charge conductivity. SHC and SHA measure the efficiency of a 
material at generating a spin current and, as such, the discovery of materials with large 
SHC/SHA has become a sought-after target for the development of high performing 
SHE-based devices. 
In this work we describe a theoretical framework to compute these parameters from 
{\it ab initio}.


Originally, the SHE was explained in terms of spin-dependent scattering 
at impurities, namely it was believed to be an entirely extrinsic effect~\cite{dyakonov1971current,hirsch1999spin}. However, it was later realized that 
the SHE also exists as a bulk band-structure effect in many non-magnetic materials 
with appreciable SOC (i.e., intrinsic SHE)~\cite{murakami2003dissipationless,sinova2004universal}. The current view is then that both intrinsic and extrinsic contributions affect the results of
 experiments in a complex and often not fully understood manner. Notably, even in those metals, such as Pt, where the SHE is attributed to the intrinsic contribution \cite{tanaka2008intrinsic, guo2008intrinsic, morota2011indication, isasa2015temperature, isasa2015erratum, Zhu2019}, 
the reported SHC and SHA still vary widely across different measures \cite{sinova2015spin, hoffmann2013spin} 
with the results that depend on the samples' crystalline quality and on the presence of 
unavoidable impurities. These are factors, which altogether remain difficult to characterize
\cite{sagasta2016tuning}. Moreover, it is important to bear in mind that, in experiments, 
the SHC can only be extrapolated indirectly, after rather complex fitting procedures based 
on models, since the spin currents, unlike the charge currents, are not directly observable
quantities. The SHE thus manifests itself as a surface spin accumulation, which can be probed in 
optical experiments \cite{kato2004observation,sih2005}, but can not directly be measured 
in a device. Furthermore, the SHE is generally accompanied by other charge-to-spin conversion
phenomena \cite{Otani2017}, most notably the so-called Rashba-Edelstein or inverse 
spin-galvanic effect (ISGE) \cite{ga.tr.19,ed.90}. This generates an additional spin-polarization 
of the conduction electrons induced by the charge current. Separating such current-induced 
spin-polarization from the spin accumulation due to the SHE remains a challenging task and 
a debated problem \cite{al.ma.15,du.ga.20,sh.fe.21,yu.li.18,dr.to.23}.

Withstanding these practical as well as fundamental issues, {\it ab-initio} calculations 
have taken a significant role in the study of the SHE 
\cite{lowitzer2011extrinsic,fedorov2013analysis,chadova2015separation,wa.we.16,honemann2019spin,nair2021spin}. Most works to date rely on two complementary linear-response transport approaches, 
namely the Boltzmann equation and the Kubo formalism, both using the electronic structure 
obtained from Kohn-Sham (KS) density functional theory (DFT) \cite{kohn.99}. 
The Boltzmann transport equation is employed for elemental metals with point impurities 
and thermal disorder, and gives SHC values in the diffusive transport regime. In contrast, 
the Kubo formalism is generally applied
for calculating the intrinsic SHC of bulk materials in terms of the Berry curvature of the 
energy bands \cite{PhysRevLett.95.156601, guo2008intrinsic, guo2009ab, gradhand2012first,kodderitzsch2015linear,ryoo2019computation, zhang2021different}. 

Despite their respective successes, the Boltzmann transport equation and the Kubo 
formalism treat opposite physical limits, so that a systematic study of the interplay 
and competition between the intrinsic and extrinsic SHE remains problematic. In this 
regard, a valuable alternative approach is scattering theory, where one studies electron- 
and spin-transport properties through a finite-sized central region placed between two 
semi-infinite current/voltage leads with well-defined in- and out-propagating electronic 
states \cite{sanvito}. 
The approach coincides with the Landauer-B\"uttiker formalism for
phase coherent transport \cite{La.57,Bu.86,Bu.88}. Yet, this can also be extended to describe diffusive transport, that occurs in systems where electrons scatter to impurities, phonons etc. and whose size is much larger than the electron mean free path. For this purpose, one simply needs to introduce disorder for cells of appropriate size, and average the results over different disorder 
realizations \cite{wesselink2019calculating}. Furthermore, effects due to interfaces, 
as present in actual devices, are naturally included \cite{wa.we.16}.
To date, scattering theory 
calculations for the SHE have used a linear-response implementation based on wave functions, 
which are matched at the boundaries between the central region and the 
leads \cite{Kh.Br.05,xi.zw.06}. The corresponding results have focused on diffusive transport, eventually recovering the temperature dependence of both the 
longitudinal conductivity and the SHC observed in experiments \cite{li.yu.15,nair2021spin2}.

Scattering theory can be implemented in a more versatile way by using the non-equilibrium 
Green's functions (NEGF) technique \cite{Datta} rather than matching wave-functions. The 
combination of NEGF with DFT is then generally called DFT+NEGF \cite{Ba.Mo.02,Ta.Gu.01}. 
On the one hand, when using wave-functions, the details of the central region's electronic 
structure are not readily available, since the current through a system is determined solely 
from the asymptotic states in the current/voltage electrodes. On the other hand, in 
DFT+NEGF, one obtains simultaneously both those electronic structure details and the 
transport coefficients, understating the relation between them. Specifically, DFT+NEGF 
enables us to describe how the density of states, the charge density, the magnetic moments, 
and other microscopic quantities are modified by presence of a flowing charge current. 
This means that, in principle, one can describe the SHE and other charge-spin conversion 
phenomena, such as the ISGE, on an equal footing. In addition to that, DFT+NEGF is also 
applicable away from the linear-response regime and can be systematically combined with 
many-body methods, such as dynamical mean-field theory, to describe correlation effects 
beyond the effective single-particle DFT picture \cite{ch.mo.15,dr.ra.22}. The NEGF 
technique was applied to understand the SHE already in early model studies \cite{nikolic2005mesoscopic,ni.so.05, nikolic2006imaging}. However, to our knowledge, 
the use of the NEGF for {\it ab-initio} simulations of the SHE remains limited. 
Only very recently, a study by Belashchenko {\it et al.} \cite{Belashchenko} was dedicated 
to the SHE in Pt. 
Thus, the objective of our work is primarily to fill this knowledge gap by detailing, implementing and demonstrating the application of DFT+NEGF towards the SHE, and even more generally to other concomitant charge-to-spin conversion phenomena.

In this work, we explain how to implement and apply state-of-the-art DFT+NEGF for 
simulating the SHE and computing SHCs. We focus only on  the linear-response limit 
for a ballistic conductor, that is a conductor without disorder, 
in order to provide an assessment of the performance of our 
approach and to discuss the underlying assumptions. In particular, we note that our 
DFT+NEGF calculations entail a convenient steady-state description of the intrinsic 
SHE, which is not allowed within the Kubo formalism. It therefore requires a change 
of perspective in treating the phenomenon.

In our implementation of DFT+NEGF, charge and spin currents are computed via the 
so-called ``bond currents approach'' \cite{nikolic2006imaging, Todorov_2002, Theodonis_2006, wa.we.16, wesselink2019calculating, xie2016spin, rungger2020non, droghetti2022spin}, 
which will be explained in detail here. 
The calculations are carried out to extract the SHC of the $5d$ metals with fcc 
or bcc structure, which show isotropic SHE, meaning that the SHC 
does not depend on the crystal orientation. In particular, we discuss in great detail 
the results for Pt, which is found to have the largest SHC value, as expected 
based on both experimental and theoretical literature. We further compare our approach 
and some of our results to those contained in the recent NEGF-based work by Belashchenko 
{\it et al.} \cite{Belashchenko}, although we stress that our interest is in the systems without disorder, while those authors studied the disordered systems.
Finally, we demonstrate that DFT+NEGF can also describe current-induced 
modifications of the spin density. The ISGE, understood as global spin-polarization 
of the conduction electrons, is forbidden by symmetry in bcc and fcc materials. Nonetheless, we predict that a spin dipole moment appears over the unit cell of these materials as the charge current breaks time-reversal invariance. 

In summary, the paper consists of three main parts. In the first, we lay down the extension of DFT+NEGF to the SHE. To the best of our knowledge, this has never been thoroughly presented before in any previous study in literature. In the second, we provide accurate SHC values for $5d$ metals in the absence of disorder, expanding upon the existing literature, which primarily focuses on the diffusive regime. Lastly, we predict the current-induced spin dipole as an effect to be added to the list of charge-spin conversion phenomena.

The paper is organized as follows. In Section~\ref{sec.theory}, we describe the 
DFT+NEGF formalism, reviewing the method (Section \ref{sec:level2}) and 
introducing the bond currents approach (Section~\ref{sec.bond_currents}). In Section~\ref{sec:level3}, we provide the computational details of the calculations and in Section~\ref{sec:level4}, we present our results. In particular, we first describe the 
calculations for Pt and then those for other fcc and bcc $5d$ metals. In Section~\ref{ssec:num43}, we analyse the current-induced spin dipole. Finally, we conclude in Section~\ref{sec:level6}.  

\section{Theoretical framework} \label{sec.theory}
We describe here how DFT+NEGF is applied to study the intrinsic SHE. The problem is 
intuitively rather simple, as one just has to compute longitudinal charge currents and 
transverse spin currents in a conductor by using the 
standard implementation of the method. However, some care is needed at the conceptual 
level. In fact, our approach relies on a physical picture rather different from that 
underlying the description of the SHE within the Kubo formalism. As such, we start this 
section by presenting a few qualitative considerations, which help to understand the 
associated change of perspective. We leave the presentation of the DFT+NEGF equations 
for the second part of the section.  
\begin{figure}[t]
\includegraphics[width=0.5\textwidth,clip=true]{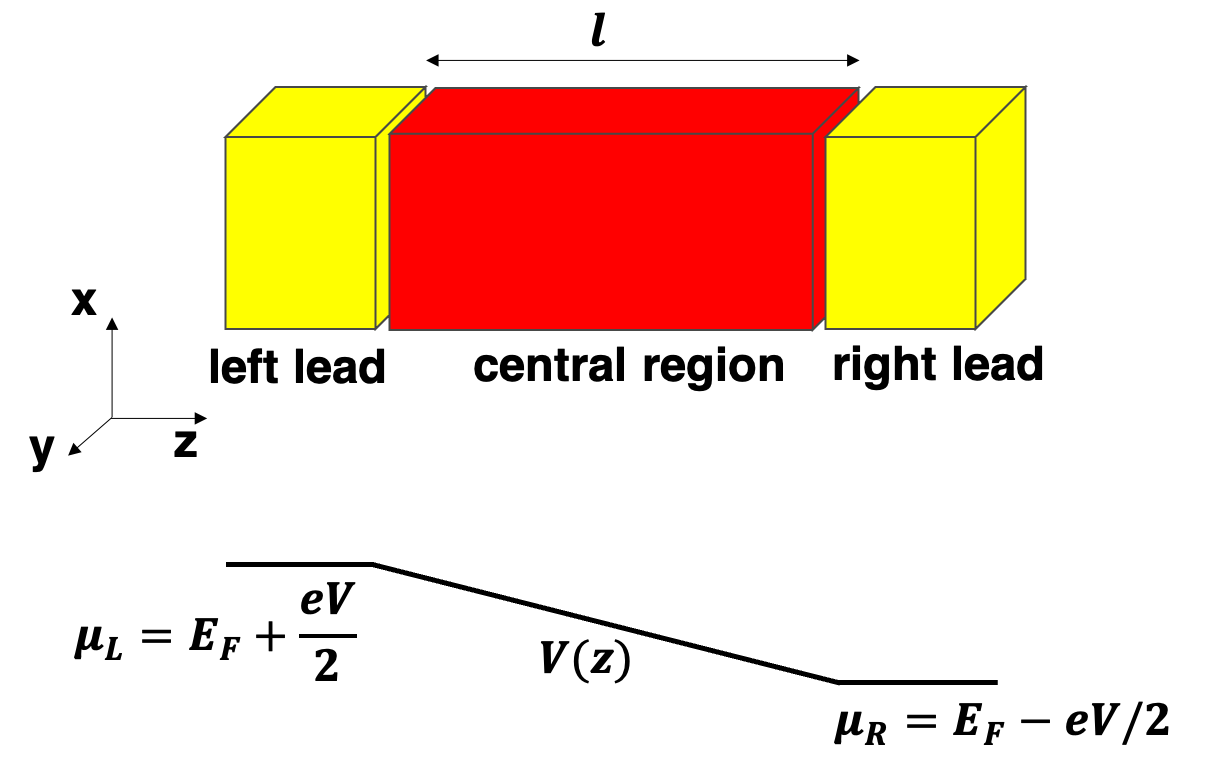}
\caption{\label{fig:figure-scattering} Schematic representation of the system studied 
in a DFT+NEGF calculation. A central region (red) is connected to two leads (yellow). The bottom panel displays the electrostatic potential 
across the system under an applied bias voltage, $V$. Here $\mu_{\mathrm{L}(\mathrm{R})}$ is the chemical
potential of the L (R) lead and $E_\mathrm{F}$ is the Fermi energy.}
\end{figure}

\subsection{Intrinsic SHE in the steady state}\label{sec.physical_picture}

In the study of the intrinsic SHE based on the Kubo's formalism, one applies an external 
uniform electric field to an extended system and calculates a steady-state spin Hall current 
as the linear response in the zero-frequency limit. The coefficient between the electric field 
and the transverse Hall current is interpreted as the intrinsic part of the static SHC, 
$\sigma_\mathrm{SH}$.
Therefore, strictly speaking, the application of the Kubo formalism to the study of the intrinsic SHE formally corresponds to computing a non-stationary response to a frequency-dependent electric field. The interpretation of the low-frequency limit of this response as a physical static SHC relies on some hidden assumptions. The first one is that the system is in the stationary state, which implies the presence of some random scattering, accounted for via a relaxation time, $\tau$, and that the system size is much larger than the electron mean free path. 
The second assumption is that, in spite of the finite relaxation time $\tau$, the corresponding energy scale, $\sim\hbar\tau^{-1}$, is negligible compared to the relevant band energies and thus can be ignored in the Kubo spectral sums.
As a result, one can define an intrinsic SHC in terms of the Berry curvature of the electronic bands. 

Importantly, in a ballistic conductor, that is a system without impurities or where impurities are so sparse that the electron mean free path is longer than the dimensions of the conductor itself, the conduction electrons subject to a uniform electric field are accelerated continuously, and the longitudinal response diverges. In fact, in frequency domain, the longitudinal conductivity grows as 
$\sigma\sim 1/\omega$ and thus, it diverges in the static limit. In contrast, $\sigma_\mathrm{SH}$ 
remains finite, making the intrinsic SHA, $\theta_{\textrm{SH}}=\sigma_\mathrm{SH}/\sigma$
undefined. This apparent problem reflects the specific physics of the intrinsic SHE in 
ballistic conductors, that is the spin Hall current is produced only by accelerated electrons. 
A Gedanken non-equilibrium state of an ideal conductor, in which the conduction electrons 
have reached a steady-state motion after the application of an electric field pulse, has 
zero steady-state Hall current, while it is characterized by a non-zero spin polarization
equal to the spin transferred across the unit cell during the transient time of the accelerated 
motion. In this sense the intrinsic Hall current produced by a uniform electric field in 
the infinite ballistic conductor is, by its nature, a transient phenomenon that can not be 
described consistently using a steady-state formalism. However, this conceptual problem 
does not subsist when a uniform 
electric field is applied only inside a finite region, as is the case in DFT+NEGF transport calculations.

In DFT+NEGF, the typical considered system is subdivided into three parts, as shown in 
Fig.~\ref{fig:figure-scattering}: a central region of finite length $l$, and a left-hand side 
(L) and a right-hand side (R) semi-infinite lead, from which electrons can flow in and out.
The leads are effectively electronic baths characterized by their chemical potentials, 
$\mu_\mathrm{L}$ and $\mu_\mathrm{R}$, and temperatures, $T$. 
A bias voltage, $V$, can be applied across the central region by displacing the leads' 
chemical potentials in such a way that $eV= \mu_\mathrm{L}-\mu_\mathrm{R}$ ($e$ is the 
electron charge). The linear response limit is then achieved by taking $V \rightarrow 0$. 
Specifically, in this paper, we introduce a linear drop of the potential between the electrodes to obtain the SHE.
From a physical standpoint, such a potential drop can be attributed to the implicit assumption of a few defects within the system, which however, we consider as sufficiently sparse that the transport remains ballistic. In this picture, the accelerating electric field is then concentrated  inside the central 
region, namely over a distance $l$, yielding a well defined steady state with a finite 
longitudinal charge current $I=GV$, where $G$ is the system's conductance. Furthermore, as 
the electrons move with acceleration between the leads, the spin Hall current, $I_\mathrm{SH}$, transverse to the charge current, is also finite 
and proportional to the voltage. This means that $I_\mathrm{SH}=G_\mathrm{SH}V$, where
$G_\mathrm{SH}$ is the spin Hall conductance. In the limit of large $l$, the spin Hall 
current saturates to a fixed value and $G_\mathrm{SH}\propto\sigma_\mathrm{SH}$, where 
$\sigma_\mathrm{SH}$ is the static SHC of an infinite crystal, which can now be calculated 
within a fully consistent steady state theory. 

The SHC obtained by means of DFT+NEGF for ballistic systems does not necessarily have 
the same value as the one calculated from the Kubo formalism. This is because of the 
different setups, different assumptions and implied limits, and the different ways of maintaining the constant electric field. In 
this sense, the intrinsic SHE calculated for a ballistic system within DFT+NEGF is a 
distinct manifestation of the SHE. 

A physically insightful view of the SHC, as treated within different limits, can be obtained by drawing a parallelism with experiments. The ballistic limit will correspond to an idealized system with no or a few impurities. This is the limit considered in our DFT+NEGF calculations here. The ``clean sample'' limit corresponds to a system with a few impurities, but already behaving as described by the Drude picture of charge transport. This is the limit where the SHE is treated by means of the Boltzmann transport equation. The SHC is dominated by the extrinsic contribution due to skew (Mott) scattering, which is inversely proportional to the number of impurities. Finally, when the impurity concentration is further increased, experiments finds that the Drude charge conductivity decreases accordingly, while the SHC eventually grows and saturates to a value independent of the impurity concentration \cite{sagasta2016tuning}. This is the ``dirty sample'' limit. The extrapolated SHC value is interpreted as the intrinsic SHC, and is compared to the SHC calculated from the Kubo formalism. In other words, the intrinsic SHC, as implied by the Kubo formalism, appears for samples with high impurity concentrations. 
In contrast, the SHC reported here by us is associated to the no-impurity limit. In the dirty sample limit, the intrinsic SHE can be physically seen as due to electrons that are accelerated along the mean-free path between impurities. In the ballistic limit, the intrinsic SHE is due to electrons that are accelerated in a finite size region of a system with no impurities.

Finally, it is important to emphasise that DFT+NEGF can cover all transport regimes and limits, and not just the ballistic one addressed here. In fact, DFT+NEGF stands out as the only approach that can do so. In order to treat the dirty sample limit, in principle, one simply needs to explicitly include impurities and disorder inside the central region. Eventually, the value of the SHC calculated with DFT+NEGF in this case should converge to the Kubo SHC, when the scattering region becomes longer than the mean free path. Unfortunately, however, DFT+NEGF calculations for the dirty transport limit remain challenging 
because of the computational burden arising from both the need to simulate large system sizes and the need to average over numerous disorder configurations.

In the rest of this paper and, in particular, in Section~\ref{ssec:num41}, we will 
illustrate how these general ideas are realized in practice in calculations. 

\subsection{\label{sec:level2}The DFT+NEGF method}
We use DFT+NEGF as implemented in the {\sc Smeagol} code 
\cite{rocha2006spin,rungger2008algorithm,rungger2020non}, which is interfaced with the 
DFT package {\sc Siesta} \cite{soler2002siesta}. We employ a linear combination of numerical 
atomic orbitals basis set, $\{\phi_\alpha(\mathbf{r}) \}$, according to the multiple-$\zeta$ 
scheme, where $\alpha$ is a collective index spanning the quantum numbers $(n,l,m)$ and the 
atom to which a basis orbital belongs. Since the basis orbitals are non-orthogonal, their 
overlap integral 
$\Omega_{\alpha\beta}=\int d\mathbf{r\,} \phi^*_\alpha(\mathbf{r} )\phi_\beta(\mathbf{r})$ is 
non-zero. In practice, we use strictly confined basis orbitals \cite{soler2002siesta}, 
namely orbitals that vanish beyond a certain cutoff radius, $r_\mathrm{cut}$. Thus, the 
overlap integral between two orbitals will vanish, if their distance is larger than 
$r_\mathrm{cut}$. We adopt the convention that the longitudinal charge transport direction 
is parallel to the Cartesian $z$-axis, and periodic boundary conditions are applied in the 
transverse $xy$ plane. We then indicate with $\mathbf{k}=(k_x,k_y)$ the wave number in 
the corresponding 2-dimensional Brillouin zone. We assume that the central region is contained 
in a rectangular supercell. Although this is not strictly necessary, in practice it greatly
simplifies the analysis of the results and the calculation of the currents for the considered
materials.

Mathematically, the subdivision of a system as presented in Fig. \ref{fig:figure-scattering} 
is achieved by describing the central region in terms of its KS Hamiltonian and overlap matrices, 
$H(\mathbf{k})$ and $\Omega(\mathbf{k})$, and by representing the leads through energy- and 
wave number-dependent
complex self-energies $\Sigma_\mathrm{L, R}(\mathbf{k},E)$. The self-energies are computed by 
using the algorithm of Ref.~\cite{rungger2008algorithm} and using the leads' Hamiltonian 
and overlap matrices from a DFT calculation for the corresponding bulk material.

The system's electronic and transport properties are calculated by solving the NEGF 
equations \cite{Datta}. Specifically, we obtain the retarded Green’s function of the 
central region by direct inversion of the Hamiltonian with the leads' self-energies
\begin{equation}
g( \mathbf{k},E)= 
[(E+i \eta)\Omega(\mathbf{k})-H(\mathbf{k}) -\Sigma_\mathrm{L}(\mathbf{k},E)-\Sigma_\mathrm{R}(\mathbf{k},E)]^{-1},\label{retardedGF}
\end{equation}
while the lesser Green's function is defined as 
\begin{equation}
 g^{<}(\mathbf{k}, E)=
i\, g(\mathbf{k},E)[f_\mathrm{L}(E)\Gamma_\mathrm{L}(\mathbf{k},E)+f_\mathrm{R}(E)\Gamma_\mathrm{R}(\mathbf{k},E)]g(\mathbf{k}, E)^\dagger.
\label{lesserGF}
\end{equation} 
The matrices
\begin{equation}
\Gamma_\mathrm{L(R)}(\mathbf{k},E)=i[\Sigma_\mathrm{L(R)}(\mathbf{k},E)-\Sigma_\mathrm{L(R)}(\mathbf{k},E)^\dagger]
\end{equation}
represent the strength of the electronic coupling between the left (right) lead and the 
central region, $f_\mathrm{L(R)}(E)$ is the left (right) lead's Fermi function at the 
chemical potential $\mu_\mathrm{L(R)}$, and $\eta$ in Eq. (\ref{retardedGF}) is a small 
positive real number.

The density matrix of the central region is obtained by integrating the lesser Green's function over the energy
\begin{equation}
 \rho(\mathbf{k})=\frac{1}{2\pi i}\int_{-\infty}^{\infty} dE\; g^{<}(\mathbf{k},E).\label{eq.DFT_rho}
 \end{equation}
In addition, we also introduce the so-called energy density 
matrix \cite{rungger2020non,Zhang2011},
\begin{equation}
 \epsilon(\mathbf{k})=\frac{1}{2\pi i}\int_{-\infty}^{\infty}  dE\; E \, g^{<}(\mathbf{k},E),\label{eq.DFT_E_rho}
 \end{equation}
which is needed to compute the charge and spin currents as shown in the next section. 

Since $H(\mathbf{k})$ is the KS Hamiltonian, which depends itself on the charge density, 
Eqs. (\ref{retardedGF}), (\ref{lesserGF}) and (\ref{eq.DFT_rho}) need to be evaluated 
self-consistently \cite{rocha2006spin}. We note that the lead self-energies, Green's functions 
and, therefore, also the (energy) density matrix depend on the bias voltage, although we have not explicitly indicated that dependence in the equations 
above to 
keep the notation short. 

A system is in thermodynamic equilibrium with Fermi energy $E_\mathrm{F}$ when the leads 
have the same chemical potential $\mu_\mathrm{L}=\mu_\mathrm{R}\equiv E_\mathrm{F}$ 
and the same temperature. Then, the effect of applying a finite bias voltage is simulated 
by shifting the leads relative chemical potentials, as discussed in 
Section~\ref{sec.physical_picture}. This, in addition to the necessary condition of local 
charge neutrality, results in a relative displacement of the whole leads' band structure. 
In practice, we set $\mu_\mathrm{L/R}=E_\mathrm{F}\pm eV/2$ and 
$\Sigma_\mathrm{L,R}(\mathbf{k},E\pm eV/2,V)=\Sigma_\mathrm{L,R}(\mathbf{k},E,V=0)$. With 
these boundary conditions, the Hartree potential of the central region is modified to capture 
the voltage drop and can be evaluated self-consistently. However, in practice, for clean metals 
like the systems studied in this work, a self-consistent solution at finite bias can not be 
found as, physically, the electronic screening and the absence of scattering prevent a 
potential drop. We then perform self-consistent calculations only at zero-bias voltage, while 
finite-bias calculations are done non-self-consistently, taking the zero-bias density matrix 
as input, and updating it only once within the rigid shift approximation \cite{Rudnev_Sci_Adv2017,xie2016spin}. As shown in Fig. \ref{fig:figure-scattering}, the shift, 
$\pm eV/2$, applied to the leads' chemical potentials and band structure, is bridged by a 
linear ramp electrostatic potential $V(z)=-eVz/l+eV/2$ in the central region, where 
$z=0$ $(z=l)$ is the position of interface between central region and the left (right) lead. 
Thus, we effectively introduce an accelerating constant electric field inside the central 
region, and we will expect to observe a finite spin Hall current. In this paper, $V$ is 
carefully chosen to ensure that the calculations are performed in the linear-response limit 
as implied by the Landauer-B\"uttiker picture.

A system can also be driven out-of-equilibrium without applying any bias voltage, but by 
setting the leads at different temperatures, $T_\mathrm{L}$ and $T_\mathrm{R}$.
In that case, one first computes the lead self-energies using the Hamiltonians and overlap 
matrices from a DFT calculation for the corresponding bulk material with electronic temperature
$T_\mathrm{L(R)}$. Then, to obtain the density matrix, one evaluates the lesser Green's function 
in Eq.~(\ref{lesserGF}) by entering the two different temperatures $T_\mathrm{L}$ and 
$T_\mathrm{R}$ in the Fermi functions $f_\mathrm{L}(E)$ and $f_\mathrm{R}(E)$ with the same 
chemical potential $\mu_\mathrm{L}=\mu_\mathrm{R}=E_\mathrm{F}$. The temperature difference 
will result in a longitudinal thermoelectric current. However, we will have no accelerating 
electric field across the central region, and we expect no spin Hall currents. Thus, 
comparing calculations for systems driven by a bias voltage and by a temperature difference 
between the leads may be useful to establish the consistency of the proposed picture for the 
SHE.

Since we have to include the SOC in our calculations of the SHE, the Hamiltonian of the 
central region assumes a non-collinear form, and it is composed of the $2\times2$ blocks 
$H_{\alpha\beta}(\mathbf{k})=H^0_{\alpha\beta}(\mathbf{k})\mathbb{1}+\mathbf{H}_{\alpha\beta}
(\mathbf{k})\cdot \pmb{\sigma}$ for any pair of basis orbitals with indices $\alpha$ and $\beta$.
Here, $H^0_{\alpha\beta}(\mathbf{k})$ and $\mathbf{H}_{\alpha\beta}(\mathbf{k})=[H^x_{\alpha\beta}(\mathbf{k}),H^y_{\alpha\beta}(\mathbf{k}),H^z_{\alpha\beta}(\mathbf{k})]$ describe the 
spin-independent and spin-dependent parts of the Hamiltonian matrix elements, respectively. 
$\mathbb{1}$ is the $2\times2$ identity matrix, and $\pmb{\sigma}=(\sigma^x,\sigma^y,\sigma^z)$ 
is a vector of Pauli matrices. Similar expansions are written also for the density matrix, 
$\rho_{\alpha\beta}(\mathbf{k})=\rho^0_{\alpha\beta}(\mathbf{k})\mathbb{1}+\pmb{\rho}_{\alpha\beta}(\mathbf{k})\cdot \pmb{\sigma}$, and the energy density matrix, 
$\epsilon_{\alpha\beta}(\mathbf{k})=\epsilon^0_{\alpha\beta}(\mathbf{k})\mathbb{1}+\pmb{\epsilon}_{\alpha\beta}(\mathbf{k})\cdot \pmb{\sigma}$. In contrast, the overlap matrix 
is spin independent, that is $\Omega_{\alpha\beta}(\mathbf{k})=\Omega^0_{\alpha\beta}(\mathbf{k})\mathbb{1}$.

An important aspect of DFT+NEGF as presented here is that, in addition to enabling the 
calculation of the transport properties (as shown below), it also gives access to the 
electronic structure of the central region. In particular, using the spin-dependent part of 
the density matrix, we can compute  the spin density at any position $\mathbf{r}$ in space
\begin{equation}
\mathbf{s}(\mathbf{r})=\frac{1}{N_{\mathrm{BZ}}}\sum_{\mathbf{k}}w_{\mathbf{k}}\sum_{\alpha,\beta} \phi^{*}_\beta(\mathbf{r}) \pmb{\rho}_{\alpha\beta}(\mathbf{k}) \phi_\alpha(\mathbf{r}),\label{eq.spin}
\end{equation}
where $N_{\mathrm{BZ}}$ is the number of wave numbers (i.e, $\mathbf{k}$ points) in the transverse Brillouin zone, 
and $w_\mathbf{k}$ is the weight associated to each of them. Such spin density is calculated 
in Section~\ref{ssec:num43} showing that the SHE is accompanied by a current-induced spin 
dipole. 

\begin{figure}[ht]
\includegraphics[width=0.4\textwidth,clip=true]{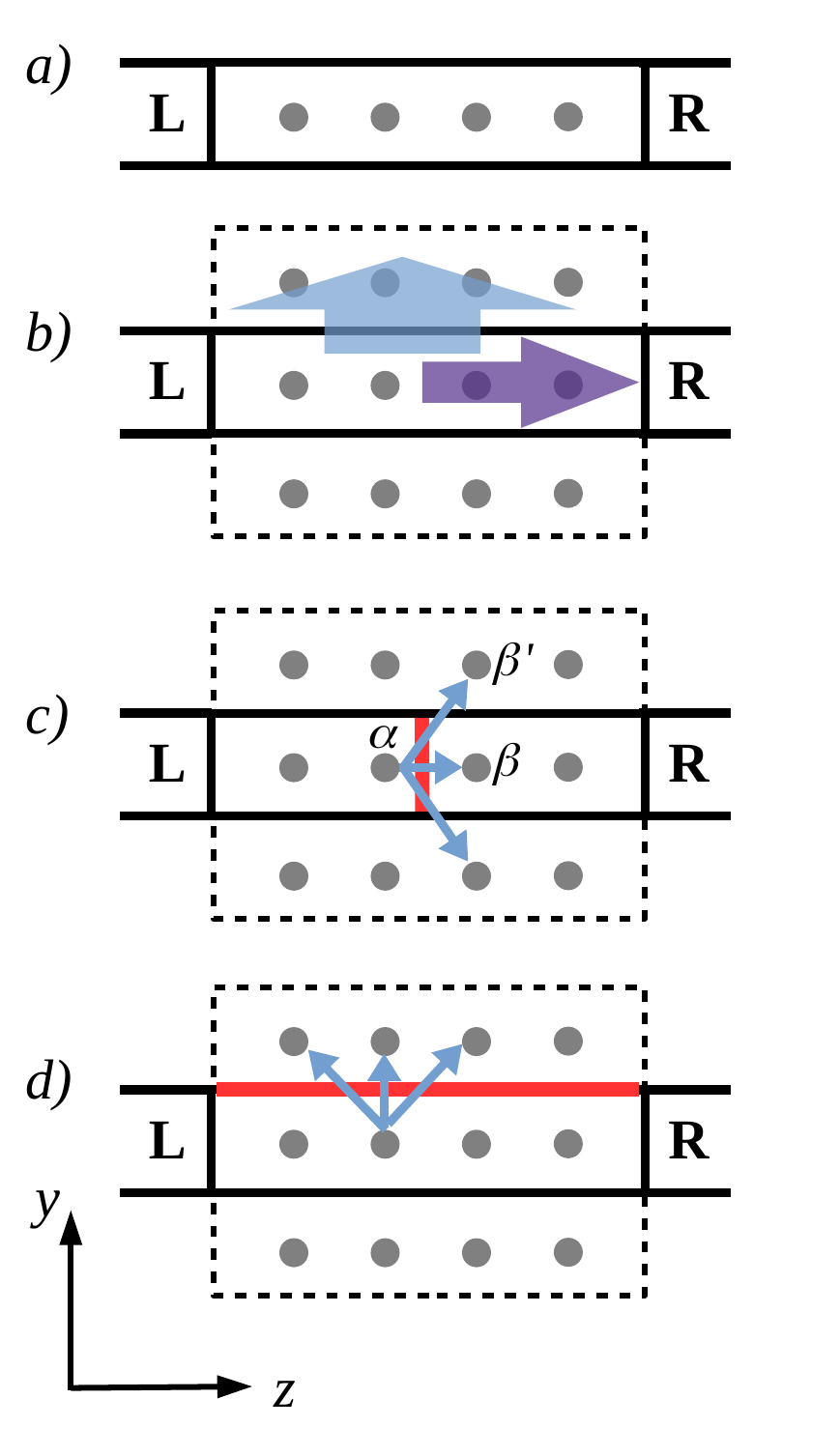}
\caption{\label{fig:figure-bond_currents} 
Scheme illustrating the calculation of the 
total current via a summation of bond currents. (a) We consider a model system with a 
supercell containing four single-orbital atoms, and we assume that the bond currents 
are non-zero only for nearest neighbour orbitals. (b) The longitudinal charge current (purple arrow) is the current flowing from the left to the right lead. The spin Hall current (cyan arrow) is spin current directed from the central region supercell to its periodic replica in the transverse direction. (c) 
The longitudinal current is calculated by
summing all the bond currents through a cross section plane of the central region and connecting the orbitals on one side of such cross-section 
plane to the orbitals on the other side of it. In such a summation, we must include bond currents, 
$\mathcalI{I}_{\alpha\beta}$, between orbitals, which are both inside the supercell central 
region, as well as the bond currents $\mathcalI{I}_{\alpha\beta'}$, which connect an orbital 
inside that supercell to an orbital outside it and centered at the supercell periodic replica. 
(d) The total transverse spin current is obtained by summing all the bond currents connecting the 
orbitals inside the central region supercell to the orbitals in the replicated supercell.
The periodic replicas of the supercell are here depicted as dashed boxes.}
\end{figure}

\subsection{Bond currents for SHE calculations}\label{sec.bond_currents}

The SHE is studied by computing the transverse spin Hall current that emerges when a bias voltage is 
applied across a system, together with the associated longitudinal charge current. 
In our set-up, which displays periodic boundary conditions in the transverse directions, the spin Hall current can be seen as a circulating current, exiting the central region supercell from one side and re-entering from the opposite side along one of the transverse directions. In practice, such spin-Hall current is calculated in terms of the spin currents directed from the central region supercell to its periodic replica in the transverse direction, as shown in Fig. \ref{fig:figure-bond_currents}b and further described in the following.

We use the bond-current approach \cite{Todorov_2002}, which has lately found wide application to the study of charge transport in systems ranging from molecular junctions to graphene nano-ribbons (see, for example, Refs. \cite{Solomon2010,ca.ga.19}). The approach was generalized to obtain spin currents first within the tight-binding formalism \cite{nikolic2006imaging, Theodonis_2006} and, more recently, within Kohn-Sham DFT  \cite{wesselink2019calculating, rungger2020non,droghetti2022spin}. 
Although the calculations presented in this paper are performed with the {\sc Smeagol} code, 
the equations below are general and can be applied within any standard DFT+NEGF implementation.

The current between any pair of basis orbitals, $\alpha$ and $\beta$, and associated to a state with wave-number $\mathbf{k}$ is defined as
\begin{equation}
\tilde{\mathcalI{I}}_{\alpha\beta}(\mathbf{k})=\tilde{\mathcalI{I}}^0_{\alpha\beta}(\mathbf{k})\mathbb{1}+\tilde{\mathbfcal{I}}_{\alpha\beta}(\mathbf{k})\cdot \pmb{\sigma},
\end{equation}
where 
%
\begin{equation}\begin{split}
&\tilde{\mathcalI{I}}^0_{\alpha\beta}(\mathbf{k})=\\
&\frac{4e}{\hbar}\mathrm{Im}{\Big[H^0_{\alpha\beta}(\mathbf{k})\rho^0_{\beta\alpha}(\mathbf{k}) +\mathbf{H}_{\alpha\beta}(\mathbf{k})\cdot \pmb{\rho}_{\beta\alpha}(\mathbf{k})  -\Omega^0_{\alpha\beta}(\mathbf{k})\epsilon^0_{\beta\alpha}(\mathbf{k})\Big]}\label{eq.bond_current_c}
\end{split}
\end{equation}
is the charge component and
\begin{equation}\begin{split}
&\tilde{\mathbfcal{I}}_{\alpha\beta}(\mathbf{k})=\\
&\frac{4e}{\hbar}\mathrm{Im}{\Big[\mathbf{H}_{\alpha\beta}(\mathbf{k})\rho^0_{\beta\alpha}(\mathbf{k}) +H^0_{\alpha\beta}(\mathbf{k}) \pmb{\rho}_{\beta\alpha}(\mathbf{k}) -\Omega^0_{\alpha\beta}(\mathbf{k})\pmb{\epsilon}_{\beta\alpha} (\mathbf{k})\Big]}.\label{eq.bond_current_s}\end{split}
\end{equation}
is the spin component \cite{rungger2020non,droghetti2022spin}.
The total bond current $\mathcalI{I}_{\alpha\beta}=\mathcalI{I}^0_{\alpha\beta}\mathbb{1}+\mathbfcal{I}_{\alpha\beta}\cdot \pmb{\sigma}$ between $\alpha$ and $\beta$ is then obtained by performing the summation over $\mathbf{k}$ 
\begin{equation}
\mathcalI{I}_{\alpha\beta}=\frac{1}{N_\mathrm{BZ}}\sum_\mathbf{k} w_{\mathbf{k}} \tilde{\mathcalI{I}}_{\alpha\beta}(\mathbf{k}).\label{eq.ksum1}
\end{equation}
In Eq.~(\ref{eq.ksum1}), $\mathcalI{I}^0_{\alpha\beta}\mathbb{1}$ is the charge bond current, while $\mathbfcal{I}_{\alpha\beta}=(\mathcalI{I}^x_{\alpha\beta},\mathcalI{I}^y_{\alpha\beta}, \mathcalI{I}^z_{\alpha\beta})$ is the spin bond current with the superscript indicating the spin-polarization direction.
We note that both indices $\alpha$ and $\beta$ in the equations above refer to orbitals inside the central 
region supercell. However, as already anticipated, we generally need also the bond currents that flow outside 
that supercell (see Fig.~\ref{fig:figure-bond_currents} and also appendix A in 
Ref.~\cite{droghetti2022spin}). These are obtained by considering the periodic replicas 
of the supercell, including the appropriate Bloch phase factor in the $\mathbf{k}$ summation
\begin{equation}
\mathcalI{I}_{\alpha\beta'}=\frac{1}{N_\mathrm{BZ}}\sum_\mathbf{k} w_{\mathbf{k}} e^{i\mathbf{k}\cdot \mathbf{R}_{\beta'}}\tilde{\mathcalI{I}}_{\alpha\beta}(\mathbf{k}).\label{eq.ksum2}
\end{equation}
Here $\alpha$ refers to an orbital in the central region, while $\beta'$ indicates an orbital 
located in a periodic replica of the central region supercell, equivalent to the the orbital 
$\beta$, from which it is separated by a translation vector $\mathbf{R}_{\beta'}$.

Since we use confined basis orbitals, a bond current $\mathcalI{I}_{\alpha\beta'}$ is zero 
by definition when the distance between two orbitals $\alpha$ and $\beta'$ is larger than 
the orbitals' cutoff radii. Furthermore, since the bond currents depend on the Hamiltonian 
matrix elements, they have the largest values for pairs of orbitals centered at nearest neighbour 
and next nearest neighbour atoms, while they become negligible for pairs of distant orbitals. 
Thus, the number of relevant bond currents in the sum is practically limited. Once these are
obtained, we can finally compute the total longitudinal charge current and the transverse spin
current.

The longitudinal charge current $I$ is defined as the charge current flowing through a cross 
section of the central region perpendicular to the $z$ direction, and it is calculated as
schematically illustrated in Fig.~\ref{fig:figure-bond_currents}c. We sum the charge components 
of the bond currents from the orbitals $\alpha$ on the left-hand side of the cross-section 
plane to the orbitals $\beta$ on the right hand side of it
\begin{equation}
I=\sum_{\alpha, \beta(')} \mathcalI{I}^0_{\alpha \beta(')}.\label{eq.current}
\end{equation}
Importantly, for this summation to be correctly carried out, $\beta(')$ has to run over 
orbitals inside the supercell central region as well as orbitals outside it and centered 
inside its periodic image. These two different kinds of bond currents are given respectively 
in Eq.~(\ref{eq.ksum1}) and Eq.~(\ref{eq.ksum2}). 

The spin currents can be calculated in the same way as the charge current by performing 
a summation of the bond currents, but considering the spin instead of the charge components. 
In particular, we indicate a spin current as $I^a_i$ ($i,a=x,y,z)$, where $a$ is the index 
labelling the spin-polarization direction, while $i$ indicates the flow direction along one 
of the Cartesian axis. Importantly, from a mathematical point of view, the spin currents 
$I^a_i$ are components of a second-rank pseudotensor (see, for example, 
Ref.~\cite{droghetti2022spin}). The structure of the spin current pseudotensor is 
entirely determined by the crystal symmetry. Thus, as a first validation of the calculation, 
one can check that the results for a material are consistent with the corresponding 
symmetry.
%

As already anticipated, in our set-up, a spin Hall current $I_\mathrm{SH}$ is 
defined as the transverse spin current flowing parallel to either the $x$ or the $y$ Cartesian axis 
and through the lateral side of the central region supercell,
\begin{equation}
I_\mathrm{SH}\equiv I^{a}_{x(y)}=\sum_{\alpha_i, \alpha_o} \mathcalI{I}^a_{\alpha_i \alpha_o}.\label{eq.spin_current}
\end{equation}
Here, $\alpha_i$ and $\alpha_o$ indicate the orbitals inside and outside the supercell of 
the central region, as illustrated in Fig.~\ref{fig:figure-bond_currents}d, and $\mathcalI{I}^a_{\alpha_i \alpha_o}$ is obtained from Eq.~(\ref{eq.ksum2}). We note that when calculating $I_\mathrm{SH}$,
some care is necessary to check the dependence of the results on the supercell's dimensions. This issue is discussed in the appendix of Ref. \cite{droghetti2022spin}, to which we refer for details.

The spin currents have the unit of an energy. However, here we have expressed them in 
the same unit as the charge current by introducing the factor $e/\hbar$ on the left-hand 
side of Eq.~(\ref{eq.bond_current_s}). In this way, we will be able to directly compare 
the relative magnitude of the longitudinal charge current and the spin Hall current.

The charge and spin conductances $G$ and $G_\mathrm{SH}$ can be fitted from the plot of the 
longitudinal charge and spin Hall currents, $I$ and $I_\mathrm{SH}$, as a function of the 
applied bias voltage, $V$. The obtained $G$ value at zero-temperature can then be compared to 
that given by the Landauer-B\"uttiker formula 
\begin{equation}
G=\frac{e^2}{h}T(E_\mathrm{F}),\label{eq.LB}
\end{equation}
where $h$ is the Planck constant and $T(E)$ is the zero-bias transmission coefficient 
\cite{datta2005quantum},
\begin{equation}
T(E)= \frac{1}{N_\mathrm{BZ}}\sum_\mathbf{k} w_\mathbf{k}\mathrm{Tr}[\Gamma_L (\mathbf{k},E) G(\mathbf{k},E)^\dagger \Gamma_R(\mathbf{k},E) G(\mathbf{k},E)]
\end{equation}
evaluated at the Fermi energy. In the linear-response limit, both 
Eq.~(\ref{eq.LB}) and the fit of the $I$-$V$ curve must return identical results. Furthermore, 
$T(E_\mathrm{F})$ must be equal to the number of transport channels in the longitudinal 
direction.

A bond current approach similar to ours was already employed in the scattering theory 
calculations of Refs.~\cite{nair2021spin,wesselink2019calculating,nair2021spin2} which were 
mentioned in the introduction. However, we note that the implementation in those works was based on wave-functions rather than Green's functions. 

The NEGF calculations by Belashchenko {\it et al.} 
\cite{Belashchenko} for the SHE in Pt did not consider bond currents. The 
spin Hall currents were instead obtained from ``spin-orbital'' torques \cite{be.ko.19}.
The idea there is that a change in a local spin-orbital torque must be compensated by a variation 
of a local spin current, and, as a consequence, it is possible to obtain one quantity from the other. The approach by Belashchenko {\it et al.} is equivalent to ours whenever spin currents in a material are not conserved. However, for defect-free and transversely periodic systems, like those 
studied here, spin currents are instead conserved, rendering spin-orbital torques null. Hence, the computation of spin-Hall currents must directly involve bond currents. From this point of view, our approach is more general than that used in previous works, and, as shown in the following, it allows for a detailed description of the SHE. 

\section{\label{sec:level3}Computational details}
DFT+NEGF calculations are performed with the local spin-density
approximation (LSDA) to the exchange-correlation functional 
\cite{ba.he.72,vosko1980} and the on-site approximation to 
the SOC \cite{fernandez2006site}. This latter was generally 
found accurate even for materials with non-trivial spin-textures 
(see \cite{ja.na.15,na.ru.12,Narayan2014}). Core electrons are 
treated using norm-conserving Troullier-Martins pseudopotentials 
\cite{troullier1991efficient,troullier1991efficient2} generated using 
the {\sc Atom} code \cite{atomcode}. The $spd$ valence electrons are
expanded using the numerical atomic orbital basis set of double-$\zeta$
plus polarization quality \cite{artacho1999linear,junquera2001numerical}. The cutoff radii of 
the basis orbitals for W and Pt are obtained from Ref. 
\cite{rivero2015systematic}, while those for Ta and Ir are the same 
as for W and Pt, respectively. In contrast, those for Au are taken 
from previous {\sc Smeagol} calculations for molecular junctions 
\cite{Rudnev_Sci_Adv2017}. For all materials, the pseudopotentials
and basis sets have been validated to closely reproduce the electronic
band structure obtained from the {\sc Quantum Espresso} plane-wave 
code v7.0 \cite{giannozzi2009quantum}.

The density matrix in Eq.~(\ref{eq.DFT_rho}) is calculated by splitting the integration of the lesser Green's function into the so-called equilibrium and non-equilibrium components \cite{rocha2006spin}. 
The equilibrium component is obtained by performing the integration along a contour in the complex energy plane \cite{rocha2006spin}.
We use 32 energy points along the semicircle and the imaginary line that form that contour, and we evaluate 16 poles. The non-equilibrium contribution is calculated by performing the integration over the real energy axis using at least 48 energy points within an integration range, which extends from $\mu_\mathrm{L}$ to $\mu_\mathrm{R}$. Since all the studied materials are non-magnetic, we initialize the magnetic moment of each atom to be zero within double precision. This is important to accurately resolve the current-induced spin-polarization with a reduced numerical noise.

Both the zero-bias and the finite bias calculations are performed with at least a 21 $\times$ 21 \textbf{k}-
point uniform grid in the transverse Brillouin zone. Using as an example the case 
of Pt, the convergence of the total transverse spin current with 
respect to the number of \textbf{k}-points is discussed in the
Supplementary Material \cite{suppmat}.

We consider the same material in both the leads and the central region to avoid any interface effects and, therefore, to capture bulk 
properties. The systems are presented in Fig. S1 of the Supplementary Material.
Ta and W have bcc crystal structure, whereas Ir, Pt and 
Au are fcc metals. We use the experimental lattice parameters in all 
calculations and set the longitudinal transport direction, $z$, along 
the [001] crystallographic direction. The Cartesian $x$ and $y$ axes 
are then oriented parallel to the [100] and [010] directions, 
respectively. 

\section{\label{sec:level4}Results and Discussion}

\subsection{\label{ssec:num41}Intrinsic SHE in Pt}

\begin{figure}[h]
\includegraphics{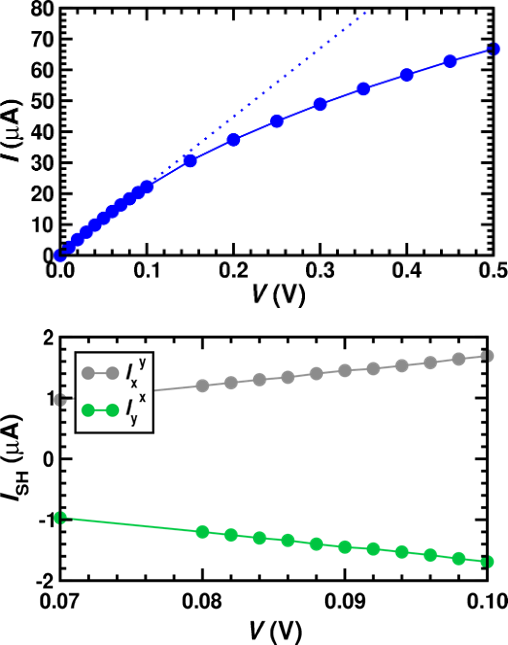}
\caption{\label{fig:figure-2} SHE computed for fcc Pt. Top panel:
longitudinal charge current, $I$, as a function of the applied bias
voltage, $V$. The circles denote the computed values, while the solid 
line is a guide to the eyes. The dotted line is the linear fit up to 
a bias value of 0.1 V. Bottom panel: transverse spin Hall currents,
$I_{\text{SH}}=I_x^y=-I_y^x$, as a function of the applied bias voltage
$V$ in the 0.07-0.10~V range. The solid line is a guide to the eyes. 
All the results are for a Pt 10-layer thick central region.}
\end{figure}

We consider here fcc Pt as an example to illustrate in detail the use of DFT+NEGF to study the SHE. The device studied, comprising the Pt leads and the central region, is shown in Fig.~S1(a) of the Supplementary Material. 
We initially consider 10 atomic
layers in the central region across which we apply different voltages, as schematically shown in Fig. \ref{fig:figure-scattering}. The results are presented in Fig. \ref{fig:figure-2}.

We find that the longitudinal charge current, $I$, is accompanied 
by transverse spin currents with perpendicular spin polarization, 
$I_x^y$ and $I_y^x$, with $I_x^y=-I_y^x \equiv I_\mathrm{SH}$. These 
are the spin Hall currents. All the other spin currents $I^a_i$,
with $a\neq x(y)$ and $j\neq y(x)$, vanish as dictated by the material's
cubic symmetry. Intuitively, a positive sign is found when the current and the spin are both aligned along the positive Cartesian directions. In contrast, the negative sign means that the current and the spin are aligned in opposite directions along the Cartesian axis, with one aligned along the positive direction and the other one along the negative direction.
Both the charge current and the spin Hall
current are seen to vary linearly as a function of the applied bias voltage for $V\lesssim 0.1$ V, which delimits the linear response regime. The charge and spin Hall conductances obtained by fitting the data to the equations $I=GV$ and $I_\mathrm{SH}= G_\mathrm{SH}V$ are $G=7 e^2/h$ and 
$G_\mathrm{SH}=0.7 e^2/h$, respectively.
The estimated $G$ 
is in agreement with the value obtained from the Landauer-B\"uttiker
formula of Eq.~(\ref{eq.LB}), with the transmission coefficient equal 
to $7$, namely the number of open transport channels at the Fermi 
energy. 
The transverse spin currents are found to vanish at zero-bias, as 
expected since equilibrium spin currents, albeit generally predicted 
in materials \cite{Rashba2003,Tokatly2008}, are forbidden by symmetry 
in fcc crystals \cite{droghetti2022spin}. Overall, our results show how the SHE emerges in the ballistic transport setup. Both the 
longitudinal charge current and spin Hall current are finite and can 
be obtained within the same computation.

Similar transport calculations can also be carried out by applying a temperature difference instead of the bias voltage between the leads, as
described in Section~\ref{sec:level2}. On doing so, we find a finite
thermoelectric longitudinal current, which depends linearly on the 
temperature difference. However, we observe no spin Hall currents. This negative result 
directly demonstrates that the emergence of the SHE requires an electric 
field to accelerate electrons.

\begin{figure}[h]
\includegraphics{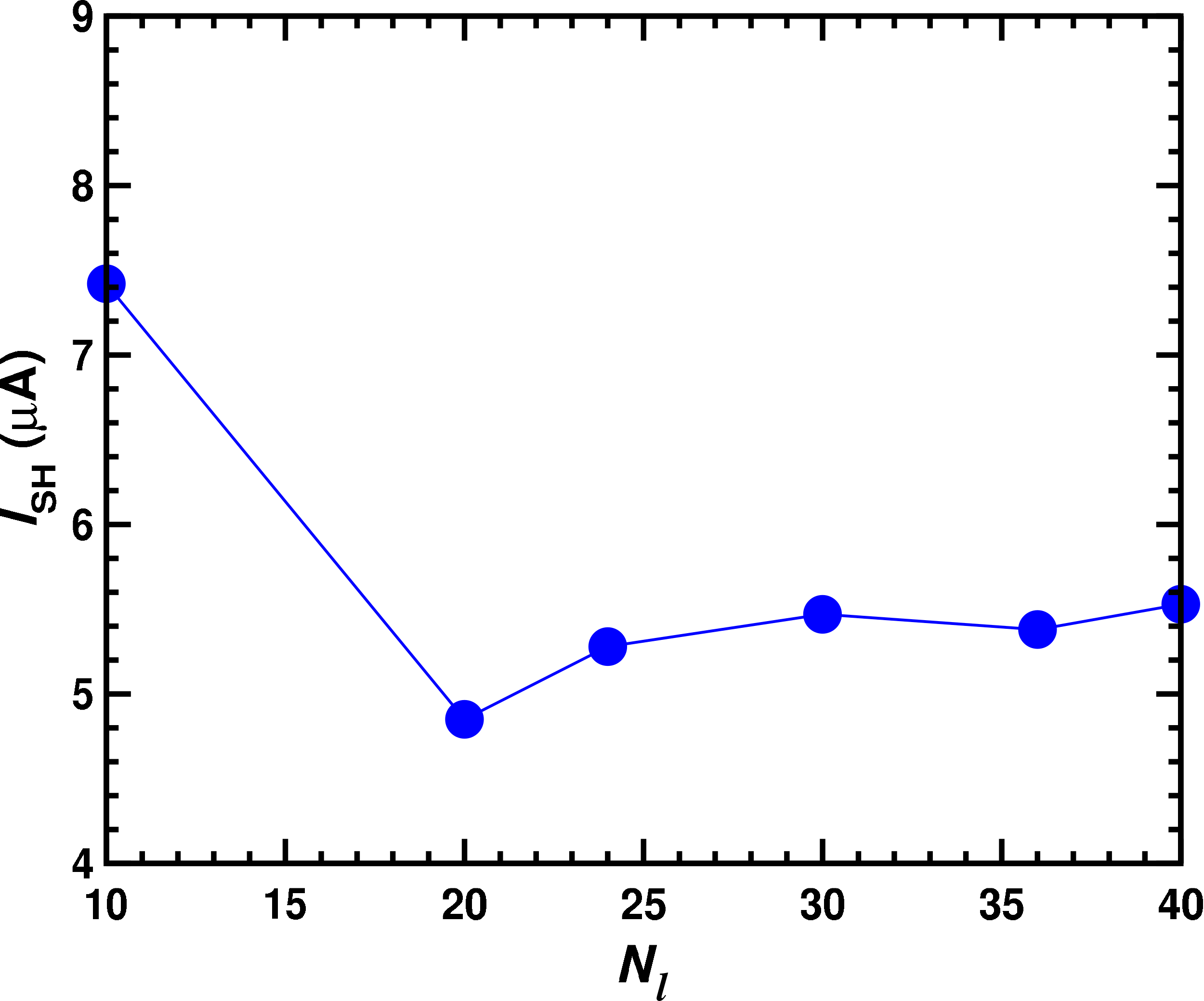}
\caption{\label{fig:figure-3} Spin Hall current, $I_\mathrm{SH}$, for 
Pt as a function of the number of atomic layers included in the 
central region. }
\end{figure}

To further analyse the SHE at the quantitative level, we now increase the length of the central region from 10 atomic 
layers ($l$=1.77 nm) to a maximum of 40 atomic layers ($l$=7.65 nm). For a constant
applied bias voltage, $V$, the longitudinal charge current does not
depend on the length of the central region. 
In contrast, the spin Hall current, $I_\mathrm{SH}$, significantly varies as a function of the number of layers, $N_l$, as shown in Fig.~\ref{fig:figure-3}.   
This different behaviour is due to the fact that our system is effectively infinite along the charge transport direction since the
leads are made of the same material as the central region. Thus, independent of the central region length, the longitudinal linear-response charge current is equal to $I=GV$ with $G$ the zero-bias Landauer-B\"uttiker conductance of the infinite system, as already mentioned above.
On the other hand, the spin Hall effect can only emerge by applying a voltage potential ramp
over the finite-sized central region of our infinite system, and the spin Hall current is confined within that region (see Fig. \ref{fig:figure-bond_currents}b). As such, it is influenced by finite size effects and hidden interfaces at the two layers, where the potential ramp starts and ends.

The largest value for $I_\mathrm{SH}$ is found for the
shortest system.
In the limit of long systems, when the electric field across the central 
region $E=eV/l$ becomes infinitely small and hidden interface effects become negligible, we reach an asymptotic value $I_\mathrm{SH}
\rightarrow \sigma_\mathrm{SH}Vd$, where $d$ is the lateral size of 
the central region. We can then estimate the SHC as argued in 
Section~\ref{sec.physical_picture}. Although the maximum number 
of layers that one can practically consider is limited by the computational cost 
of the calculations, we observe in  Fig.~\ref{fig:figure-3} that
the spin Hall current has already converged for relatively small systems. Thus, from the results of the longest studied system ($N_l=40$), we obtain that the SHC is $\sigma_{\text{SH}} \sim 1400$ 
$(\Omega \text{cm})^{-1}$ with an error of about $10\%$ due to remaining 
numerical errors in the calculation, such as those due to the 
$\textbf{k}$-point sampling and the transverse size of the supercell, analyzed in the Supplementary Material, where we present results of test calculations for different $\textbf{k}$-points and supercells.
Notably, our estimate for $\sigma_{\text{SH}}$ is within the range 
of the theoretical and experimental reports that one can 
find in literature. These go from $\sim1000$ to $\sim 2000$ $(\Omega \text{cm})^{-1}$ \cite{tanaka2008intrinsic, isasa2015temperature, isasa2015erratum, sagasta2016tuning, wa.we.16, nguyen2016spin, obstbaum2016tuning, zhang2015reduced, stamm2017magneto, guo2008intrinsic, dai2019observation}. 

Our computed SHC can also be compared to the value obtained from
standard Kubo-formalism calculations, i.e.,
$\sigma_\mathrm{SH} \sim 2200$ $(\Omega \text{cm})^{-1}$ 
\cite{guo2008intrinsic, nardelli2018paoflow}, noting that our 
result is somewhat smaller. As discussed in Section \ref{sec.physical_picture}, the DFT+NEGF and the Kubo SHCs are 
not expected to be the same, and the different results reflect the
different calculation setups and the different ways of sustaining 
the constant electric field. In this regard, it is interesting to 
note that our DFT+NEGF estimate is in quite good agreement with the value 
$\sim 1600$ $(\Omega \text{cm})^{-1}$ obtained via wave-function-based 
scattering theory calculations~\cite{wa.we.16, nair2021spin}, although
those were performed for systems including thermal disorder within 
the frozen phonon approximation.

\begin{figure}[h]
\includegraphics{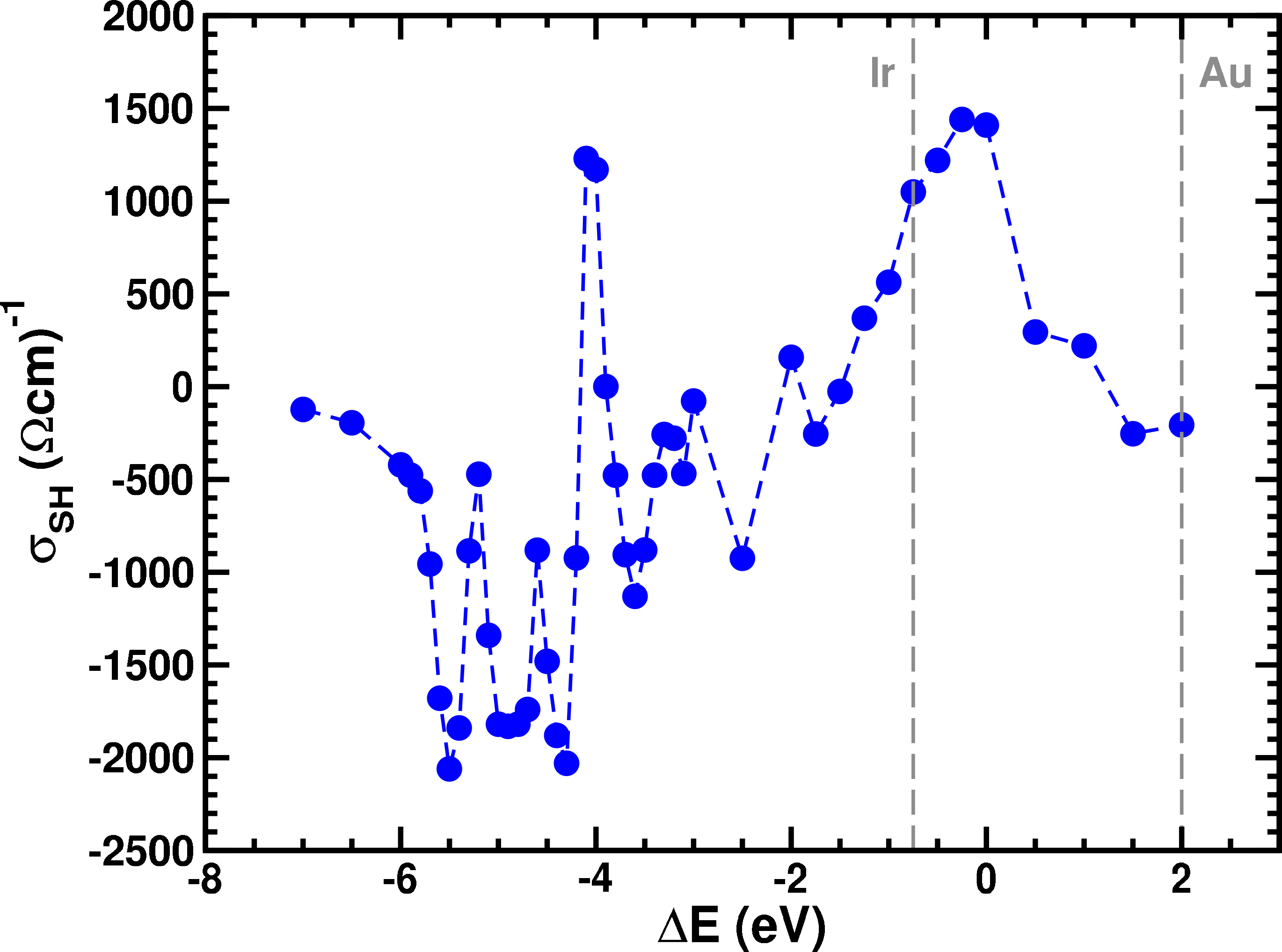}
\caption{\label{fig:figure-xyz} SHC, $\sigma_{\text{SH}}$, of 
Pt as a function of the Fermi energy position. Here $\Delta E$ is the 
position of $E_\mathrm{F}$ measured from the neutrality point, 
$\Delta E=0$ (the true system Fermi energy). The vertical gray dashed lines indicate the Fermi levels corresponding to Ir and Au.}
\end{figure}

Some physical insights into the SHE can be obtained by analysing how the SHC changes as a function of the Fermi energy, or, in other words, as a function of the $5d$ band filling. We can then compare the results 
with some generally expected trends \cite{guo2008intrinsic}. The 
calculations are practically carried out non-self-consistently by 
applying a rigid energy shift $\Delta E$ to the true Pt Fermi energy, 
$E_\mathrm{F}\rightarrow E_\mathrm{F}+\Delta E$. The results are shown 
in Fig.~\ref{fig:figure-xyz}. $\sigma_\mathrm{SH}$ has 
a broad positive maximum centered at $\Delta E=0$ eV, namely at the 
Pt true Fermi energy. It then rapidly declines as $\Delta E$ increases
and the Fermi energy exits the $5d$ band. On the opposite end, 
$\sigma_\mathrm{SH}$ is found to be negative for $\Delta E\approx$ -$4.25$ eV, 
and eventually vanishes for $\Delta E <$ -$7$ eV when the $d$ band
becomes empty. 

The SHC is generally positive (negative) when the $d$ band 
occupation is above (below) half-filling. This
qualitative behaviour is consistent with the results from the literature \cite{guo2008intrinsic,tanaka2008intrinsic,morota2011indication}
obtained with different implementations of the Kubo formalism as it
reflects the underlying electronic structure of Pt. Interestingly, the 
broad peak centered at $\Delta E=0$ eV is reminiscent of the one seen in 
the standard plot of the energy-dependent Kubo SHC reported, for example,  in
Fig. 1 of Ref. \cite{guo2008intrinsic}. 

A particularly interesting energy region in the plot of
Fig.~\ref{fig:figure-xyz} is for $\Delta E$ between -$6$ and -$2$ eV. There, the SHC has a very strong dependence on the Fermi energy position, and we can observe sharp fluctuations, not seen in the energy-dependent Kubo SHC of Ref. \cite{guo2008intrinsic}. 
This behaviour is likely related to the complex and intricate $d$ band structure around that energy region.
In particular, we note a distinct feature at 
$\Delta E \sim$ -4 eV, where the SHC suddenly switches from a negative 
to a large positive value.
Regarding this point, it is 
useful to compare our results to the ones by Belashchenko {\it et al.} 
\cite{Belashchenko}, who performed DFT+NEGF calculations for Pt in 
the presence of Anderson-like disorder. For low disorder, they observed a very similar feature to ours at $\Delta E \sim$ -4 eV in the SHC vs $\Delta E$ plot (compare Fig.~\ref{fig:figure-xyz} with Fig.~4 
in Ref. \cite{Belashchenko}). However, in their calculations, that feature disappears with
increasing disorder, and they eventually recover results comparable 
to those from the Kubo formalism. 
This finding qualitatively supports the conjecture brought forward 
in Section~\ref{sec.physical_picture} that the SHC obtained by 
DFT+NEGF should converge to the Kubo SHC in the presence of 
scattering, and specifically, when the length of the central region becomes longer than 
the electron mean free path.

Finally, we can complete our analysis by also computing the SHA,
$\Theta_{\mathrm{SH}}$, besides the SHC. In literature, the SHA is conventionally
defined as the ratio of the SHC to the longitudinal charge conductivity.
However, such a definition is valid for diffusive transport, whereas here 
we work within the Landauer-B\"uttiker picture of quantum transport, 
and, as such, there is no notion of charge conductivity. We therefore introduce 
the alternative definition, $\Theta_{\mathrm{SH}} =G_\mathrm{SH}/G$, 
namely we define the SHA in terms of the conductance. In the case of Pt, 
this ``ballistic SHA'' is then readily obtained, and we find 
$\Theta_\mathrm{SH} \sim 17 \%$ when $G_\mathrm{SH}$ is that of 
the longest system. The results can not be directly compared to 
experiments, as the ballistic conductance of Pt has never been measured. 
However, the ballistic SHA still serves as an intrinsic parameter 
uniquely relating the SHE to the material's band structure. In the following, 
it will be used to compare the SHE across different $5d$ metals.      

\subsection{\label{ssec:num42}Intrinsic SHE across cubic $5d$ metals}
We now extend our study to the other $5d$ metals with either bcc or 
fcc structures, where the SHE is isotropic. In all these materials, we find that the longitudinal charge current is accompanied by transverse
spin currents, $I_x^y = -I^x_y \equiv I_\mathrm{SH}$, like in the case of Pt. The 
SHC is estimated by considering central regions of different 
lengths, following the same approach as detailed in the previous subsection. The plots of $ I_\mathrm{SH}$ as a function of the number of atomic layers is presented in Fig. S3, S4, and S5 of the Supplementary Materials for W, Ir, and Au, respectively. The results vary from material to material as they are determined by the details of the electronic structure, which stem from the lattice and chemical composition. Nevertheless, we find that, for all considered materials, $ I_\mathrm{SH}$ has converged to within approximately $10\%$ for the longest systems that we could treat. 
 
Among the fcc materials, Pt has by far the largest SHC, $\sigma_{\text{SH}} \approx 1400$ $(\Omega \text{cm})^{-1}$, which was 
calculated
in the previous section. At the opposite extreme there is Au with a
vanishing SHC equal to about 60 $(\Omega \text{cm})^{-1}$. Ir has a somewhat intermediate SHC, $\sigma_{\text{SH}}\sim 270$ $(\Omega \text{cm})^{-1}$. 
These results are approximately the same that one would extract from 
Fig. \ref{fig:figure-xyz} at the corresponding materials' band filling, 
i.e., by assuming a rigid band shift of the Pt bands when going from 
Pt to Ir to Au, as often done in literature \cite{tanaka2008intrinsic}.

For the two bcc materials, Ta and W, we obtain similar intermediate
absolute SHC values, but with opposite signs, $\sim 660$ and 
$\sim$-$550$ $(\Omega \text{cm})^{-1}$, respectively. These values could
not be inferred from Fig.~\ref{fig:figure-xyz}, which describes the 
fcc compounds and not the bcc ones. However, the rigid 
band shift approximation is still valid as long as the proper bcc 
lattice is considered. In fact, the SHC of W can be obtained from 
the calculations for Ta by simply shifting the Fermi energy to 
artificially increase the $d$ band filling, as shown in Fig. S2 of the Supplementary Material \cite{suppmat}. Interestingly, the result 
for W is quite close to the estimate from the Kubo formalism, that 
is $765$ $(\Omega \text{cm})^{-1}$ \cite{PhysRevMaterials.4.094404}.
 
\begin{figure}[h]
\includegraphics{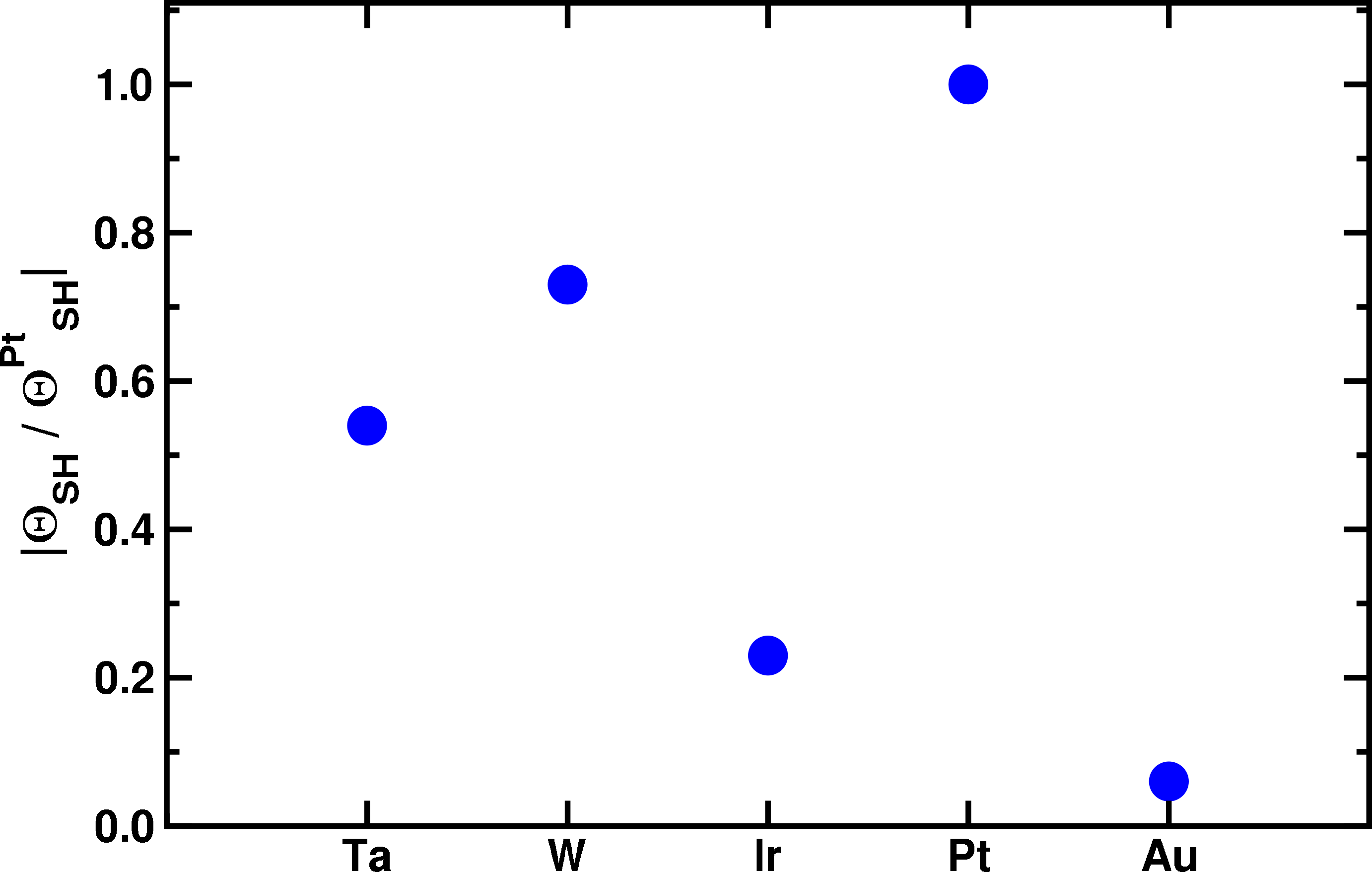}
\caption{\label{fig:figure-8} Absolute value of the SHA relative 
to that of Pt, 
$\vert\Theta_{\mathrm{SH}}/\Theta^{\mathrm{Pt}}_{\mathrm{SH}}\vert$, 
for the $5d$ metals considered in this work.}
\end{figure}

In order to compare the SHE across the various $5d$ materials, we
compute their SHA, $\Theta_{\mathrm{SH}}$, as defined in the previous
section. The results, given as a fraction of the value for Pt, 
$\vert\Theta_{\mathrm{SH}}/\Theta^{\mathrm{Pt}}_{\mathrm{SH}}\vert$, 
are displayed in Fig. \ref{fig:figure-8}. We observe a rough qualitative dependence on the partial $d$-electron 
count. The SHA has the largest 
value for Pt ($d^9$), then drops for Ir ($d^7$) and is further enhanced 
for W ($d^4$) and Ta($d^3$). Interestingly, although the SHC of Pt is 
almost three times larger than that of W, the SHAs of the two materials 
are comparable. This is because, together with the spin conductance, 
the charge conductance of W also decreases and becomes equal to 
$3e^2/h$. 

Finally, we note that W can also be found in the metastable, 
so-called, $\beta$-phase with the A15 cubic structure. Since this system  
has been reported to have a large isotropic SHE \cite{pai2012spin, hoffmann2013spin, sinova2015spin, hao2015beta, hao2015giant, qu2014self, karnad2018evidence, qu2018inverse, liu2012spin, hahn2013comparative,PhysRevMaterials.4.094404}, we have considered it as well. 
In our calculations, the SHC of $\beta$-W is $-1094$ $(\Omega \text{cm})^{-1}$,
which is about twice the value of bcc W and close to that of Pt. Our result therefore confirms the claims of its giant SHE. 

\subsection{\label{ssec:num43}Current-induced spin dipole} 

The DFT+NEGF method allows us to obtain the electronic structure of 
our systems through the same calculations performed to estimate the 
SHC. This is a clear advantage of our approach compared to the 
wave-function-based implementations of scattering theory, where 
the details of the electronic structure of the central region are
not explicitly considered. Here we analyze in particular the 
spin density, defined in Eq. (\ref{eq.spin}).

\begin{figure}[h]
\includegraphics{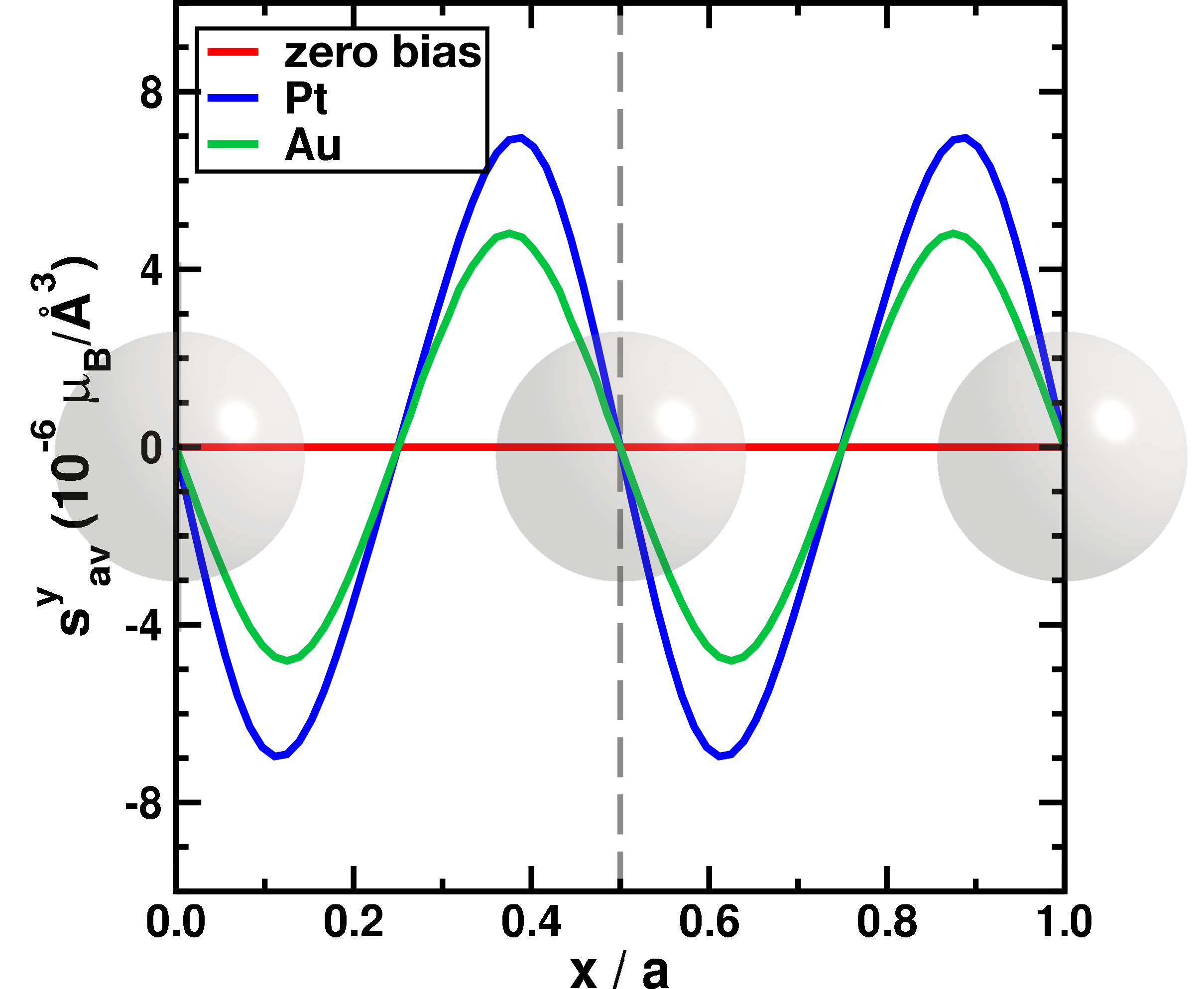}
\caption{\label{fig:figure-4} Spin-density profile $s_{\mathrm{av}}^{y}(x)$ for Pt (blue) and Au (green) as a function of the transverse $x$ coordinate in units of the lattice constant $a$. The red line indicates the average density computed at zero bias. }
\end{figure}

At zero bias, the spin density $\mathbf{s}(\mathbf{r})=[s^x(\mathbf{r}),s^y(\mathbf{r}),s^z(\mathbf{r})]$ is zero everywhere 
in the central region. In contrast, at finite bias and in the presence of a longitudinal charge
current, we find a position-dependent
modulation of the transverse spin density components, $s^x(\mathbf{r})$ and
$s^y(\mathbf{r})$. In order to practically visualize this effect, we plot in 
Fig. \ref{fig:figure-4} the so-called ``spin density profile'' 
\cite{to.kr.15, dr.to.23} in the transverse direction, namely the
spin-$y$ density averaged over the $yz$ plane, 
$s_{\mathrm{av}}^{y}(x)=(ld)^{-1}\int_0^d \int_0^l dydz\, s^{y}(x,y,z)$, 
with $l$ and $d$ the length and the lateral size of the cell. 
Similarly, we can also define the averaged spin-$x$ density 
$s_{\mathrm{av}}^{x}(y)$, which behaves in the same way as  
$s_{\mathrm{av}}^{y}(x)$. The results in Fig. \ref{fig:figure-4} are 
for Pt and Au and are obtained for the same longitudinal charge current
$I$ = 5.8 $\mu$A.

The two materials behave in a qualitatively similar way. The spin
density profile is periodic (almost sinusoidal), with a period equal to 
half of the cubic lattice parameter, i.e. $a/2$, and integrates to 
zero over the unit cell. In practice, we observe the emergence of a 
spin dipole with the positive and negative polarities centered 
between the atoms. Interestingly, although the SHE of Au is almost
negligible compared to that of Pt, we see in Fig. \ref{fig:figure-4} that the spin dipoles for the two metals are comparable.

Mathematically, the spin dipole in a crystalline material is described by the first momentum 
of the spin density and has components $
\mathcal{M}_{i}^{a}=\int_\mathcal{V} d\mathbf{r} \,\delta r_i\, s^a(\mathbf{r})$, where $\delta r_i$ is the component $i=x,y,z$ of 
the displacement vector pointing from the negative dipole polarity to 
the positive one, and $\mathcal{V}$ is the unit cell volume. Since the
spin density is an axial vector and the displacement is a polar vector, 
the spin dipole is a rank-two pseudovector, like the spin current (see
Section~\ref{sec.bond_currents}), and it has the same structure. In 
the non-magnetic materials considered here, the time-reversal symmetry
dictates that pseudotensors vanish. Therefore, at equilibrium, no spin dipole is allowed, and all the components $\mathcal{M}_i^a$ are zero. In 
contrast, when we drive a longitudinal charge current, the time-reversal
symmetry is broken and, as a consequence, a spin dipole can emerge.

The phenomenon is clearly reminiscent of the ISGE, but some care is needed when making such a connection. On the one hand, in systems hosting
the ISGE, a charge current induces a non-equilibrium spin density, which, when integrated over the entire volume,
results in a global magnetization \cite{Calavalle2022}. On the other hand, in our systems, such global magnetization
is zero. 
The ISGE is in fact only 
allowed in a subset of non-centrosymmetric materials called gyrotropic 
\cite{te.ka.23}, whereas the fcc and bcc structures are 
centrosymmetric.
In spite of that, our calculations predict that a non-zero spin 
density still appears locally in the form of a spin dipole over the unit 
cell. In other words, we may say that the ISGE is absent as a bulk unit 
cell property, but is present as a local property. This situation is 
comparable to altermagnetism, when the breaking of the time reversal 
symmetry is characterized by a local spin polarization with a vanishing 
net spin \cite{PhysRevX.12.040501}. In fact, the observed current-induced spin dipole in inversion symmetric materials can be also 
understood as a current-induced altermagnetism. If the inversion 
symmetry of one of our systems was broken making it gyrotropic, we 
would recover the global ISGE which could be viewed as a current-induced 
form of ferromagnetism.

\begin{figure}[h]
\includegraphics{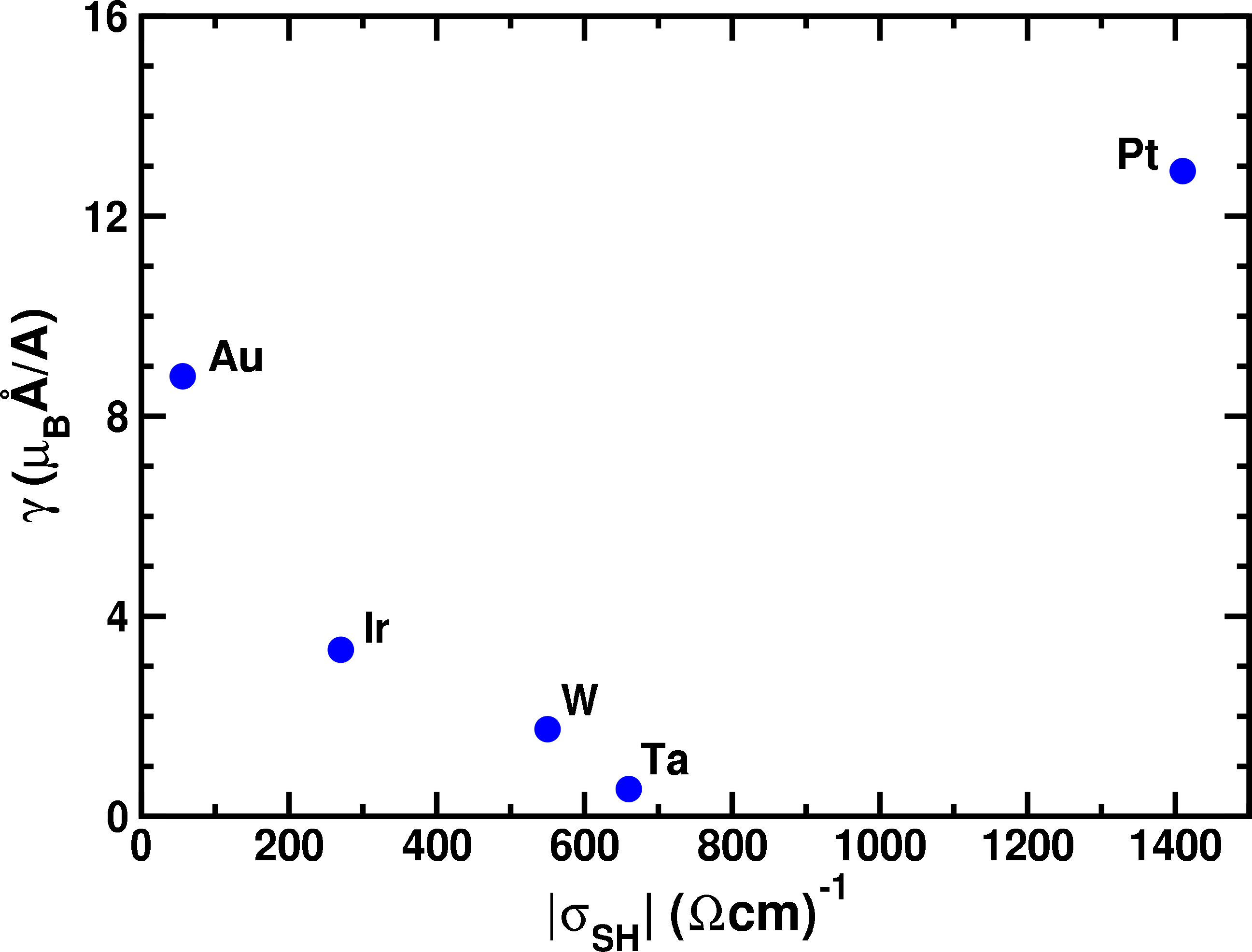}
\caption{\label{fig:figure-5} The coefficient $\gamma$ (see text) 
plotted against the absolute value of the intrinsic SHC for all the 
5$d$ metals considered in this work.}
\end{figure}

In order to perform a quantitative analysis of the current-induced 
spin dipole in our set-up, we find convenient to estimate the momentum
of the spin density profile $\mathcal{M}=\int_0^{a/2} dx\,  s_\mathrm{av}^y(x)(x-x_0)$ with $x_0$ the center of the dipole. 
Since $\mathcal{M}$ depends linearly on the longitudinal charge current 
density $j=I/d^2$, for small applied bias voltages, $V\lesssim 0.1$ V, 
we define the coefficient $\gamma=\vert \mathcal{M} /j\vert$. The 
calculated values of $\gamma$ for the different materials are plotted in 
Fig. \ref{fig:figure-5} against their SHCs.  
$\gamma$ has the largest value for Au and Pt, although the two materials
have extremely different SHCs, as already observed above. In contrast,
$\gamma$ is considerably smaller for W, Ta and Ir.
Interestingly, when we look 
not only at the magnitude but also at the sign of $\mathcal{M} /j$, we find 
that this is opposite for W and Ta, just as the SHC 
computed in Section~\ref{ssec:num42}. Overall our calculation correctly
reproduces the fact that the spin dipole and the spin current share the
same symmetry. 

A further check that the spin dipole is induced by the charge current, which breaks the time-reversal symmetry, can be carried out by applying a temperature difference, instead of a voltage bias, between the leads. In this case, we observe that the spin Hall current
disappears, since electrons are not accelerated (see Section \ref{ssec:num41}), but the spin dipole persists. Furthermore, the spin
density profile appears the same for a fixed current value, 
independently of whether that current is driven by a bias voltage or 
by a temperature difference. For instance, in the case of Pt, we compute
$\gamma \sim$ 13 $\mu_B\text{\AA}/$A in both cases.

The current-induced spin dipole in $5d$ metals was recently reported 
also in Ref. \cite{dr.to.23}, where a different approach was 
employed, treating thick slabs. In that case, it is shown that the spin 
density profile develops sharp peaks at the surface atomic layers of the slabs, where 
the spatial inversion symmetry is broken, indicating the emergence of 
the ISGE. Those peaks decay in the interior of the slabs, where the spin 
density profile eventually converges to the same bulk oscillating 
behaviour and spin dipole seen here. Overall, for all considered 
materials, the results in those slabs' interior regions are in good 
quantitative agreement with those calculated here for true bulk 
systems.

The current-induced spin dipole predicted here is probably not
measurable in experiments and will not impact any performances of 
spintronic devices.  Nonetheless, our calculations serve to illustrate 
the potential of our formalism. The DFT+NEGF method can be readily used 
to study the current-induced modulation of the spin density in real 
space. When applied to gyrotropic media rather than cubic materials, we  
expect that the approach will capture the ISGE emerging alongside the 
SHE, and it will therefore allow for a comparison of the two effects' relative 
magnitudes.

\section{\label{sec:level6}Summary and conclusion}

In this work, we have explained how to study the SHE by means of 
the DFT+NEGF approach to quantum transport. The longitudinal charge and
the transverse spin Hall currents are calculated through the so-called
bond currents, which connect atomic orbitals along different directions. 
Their mathematical definition and some details for their computational
implementation have been provided. The method is rather general and can
be used in any electronic structure code based on a linear combination
of atomic orbital basis set.

DFT+NEGF calculations have been carried out for ballistic systems in the linear response limit. The SHE emerges when we
assume a linear drop of the bias voltage potential between the two leads 
so that the electrons in the central region are accelerated by a uniform 
electric field. We then obtain a finite longitudinal charge
current as well as a finite spin Hall current proportional
to the applied bias voltage. In the limit of long central regions, the
spin Hall current saturates to a fixed value, and we can extract the 
static SHC.

SHC values have been computed for the $5d$ metals with fcc and bcc 
crystal structures. We observe that the SHC exhibits a rough dependence on
the $d$-band filling. This behavior is qualitatively consistent with
previous experimental and theoretical results. The calculated absolute
value of the SHC is the largest for Pt, drops for Ir, is 
further enhanced for W and Ta, and finally vanishes for Au,
which exhibits a fully filled $d$ band.

The SHC of a material obtained by means of DFT+NEGF in the ballistic transport regime does not
necessarily have the same value as the one calculated from the standard
Kubo formalism. 
Yet, for some of the materials considered here, such as W, we find a fair
agreement between the DFT+NEGF and Kubo results. 
Eventually, we expect that the value of the SHC obtained by DFT+NEGF 
will converge to the intrinsic Kubo SHC in the presence of scattering
and when the considered central region becomes longer than the electron mean free path. This conjecture appears to be supported by recent DFT+NEGF 
calculations for Pt including large disorder \cite{Belashchenko}. 
However, a definite proof remains challenging. As such, our future studies will be dedicated towards 
improving our implementation of DFT+NEGF to treat larger systems and including 
different kinds of disorder needed to 
systematically link the extreme limits of ballistic and highly resistive transport.

Finally, we have shown that the DFT+NEGF approach also gives access 
to the current-induced modification of the spin density. This will be 
important to describe the ISGE and other charge-to-spin conversion 
phenomena. Although the ISGE is absent as a bulk 
global property in fcc and bcc metals owing to spatial inversion symmetry, the spin density is nonetheless modulated locally. In particular, we have found that the SHE is accompanied by a spin-density dipole moment over the materials' 
unit cell. We have then estimated that the amplitude of the spin-dipole
of Pt and Au is larger than that of the other $5d$ materials.

In conclusion, we have extended the DFT+NEGF approach to the study of the SHE and, more generally, of spin-charge conversion phenomena, demonstrating the potential and versatility of the method. 
Although further extensions, in particular aimed at 
including disorder, are required to mimic experiments for $5d$ metals, 
DFT+NEGF as presented here can already be applied to any nano-material
and device, where the transport remains completely determined by the 
band structure \cite{rao2023ballistic}.

\begin{acknowledgments}
A.D. would like to thank I. Rungger and M. Stamenova for useful discussions about the bond current method. A.D. was funded by Science Foundation Ireland (SFI) and the Royal Society through the University Research Fellowship URF/R1/191769. A.B. and S.S. acknowledge the support of SFI (19/EPSRC/3605) and of the Engineering and Physical Sciences Research Council (EP/S030263/1). R.G. acknowledges the support of the European Commission FET-Open project INTERFAST (ID No. 965046). I.V.T. acknowledges support from Grupos Consolidados UPV/EHU del Gobierno Vasco (Grant No. IT1453-22) 
and from Grant No. PID2020-112811GB-I00 funded by MCIN/AEI/10.13039/501100011033. The computational resources were provided by the Trinity Centre for High-Performance Computing (TCHPC) and the Irish Centre for High-End Computing (ICHEC) facilities. A.B. and R.G. have contributed equally to this work.
\end{acknowledgments}

\bibliography{references}

\end{document}